\documentclass[aps,twocolumn,superscriptaddress,showpacs]{revtex4-1}

\usepackage{amsmath}
\usepackage{graphicx}

\begin{document}

\title{Optical non-reciprocity of cold atom Bragg mirrors in motion}

\author{S. A. R. Horsley}
\affiliation{Electromagnetic and Acoustic Materials Group, Department of Physics and Astronomy,
University of Exeter, Exeter, EX4 4QL, England.}
\email{s.horsley@exeter.ac.uk}
\author{Jin-Hui Wu}
\affiliation{College of Physics, Jilin University, Changchun, P. R. China}
\author{M. Artoni}
\affiliation{European Laboratory for Nonlinear Spectroscopy \& Istituto Nazionale di Ottica, Firenze, Italy.}
\affiliation{Department of Engineering and Information Technology \& CNR-IDASC, University of Brescia, Brescia, Italy.}
\author{G. C. La Rocca}
\affiliation{Scuola Normale Superiore \& CNISM, Pisa, Italy}
\date{\today }

\begin{abstract}
Reciprocity is fundamental to light transport and is a concept that holds also in rather complex systems. Yet, reciprocity can be switched off even in linear, isotropic and passive media by setting the material structure into motion. In highly dispersive multilayers this leads to a fairly large forward-backward asymmetry in the pulse transmission. Moreover, in multilevel systems, this transport phenomenon can be all-optically enhanced. For atomic multilayer structures made of three-level cold $^{87}$Rb
atoms, for instance, forward-backward transmission contrast around $95\%$ can be obtained already at atomic speeds in the meter per second range. The scheme we illustrate may open up avenues for optical isolation that were not previously accessible.
\end{abstract}

\maketitle

Much attention has been devoted to the development of advanced materials and composite systems to achieve optical functionalities not readily available in natural media. Such optical metamaterials can be engineered to stretch the rules that govern light propagation and light-matter interaction, potentially seeding a new paradigm in all-optical, optoelectronic and optomechanical devices. Photonic crystals and negatively refracting media are prominent instances of man-made systems the optical properties of which can be tailored to a great extent. Nevertheless, some tasks are more difficult than others. Already in the familiar process of linear reflection and transmission of light~\cite{Yeh:2005uu} it is in general hard to achieve non-reciprocity. In particular, multilayer photonic structures made with linear isotropic media with dissipation and/or gain may exhibit reflection non-reciprocity, i.e., an unbalance between the forward and backward reflectivities. There are even cases in which one of them can be made negligible and the other one can increase without limit with the sample thickness~\cite{Lin}; this occurs in the so called $PT$-symmetric media which exhibit a variety of peculiar optical properties~\cite{Longhi:2010fk, Feng:2012fk}.  Yet, it is not possible to achieve a non-reciprocal transmissivity in such linear and passive systems.  Transmission reciprocity is almost ubiquitous in optics~\cite{Yeh:2005uu,Fan}.
\par
Non-reciprocal transmission is however rather desirable for information processing and crucial to the development of optical-based functional components in photonics. In much the same way in which electrical non-reciprocity has been realized through diodes, devising an \textit{optical diode} is challenging, even in theory.  Ideally, an optical diode would allow total light transmission over a bundle of wavelengths in one direction, providing total isolation in the reverse direction. Previous work on non-reciprocal transmission has been based on either magneto-optical effects~\cite{Levy:05,Zaman:2007fk,Dtsch:2005uq} or non linear processes~\cite{Fujii:2006kx,Zhou:2006vn}. Other mechanisms have also been explored~\cite{Gevorgyan:2006uq,1994JAP....76.2023S, Feise:2005bm} and realized experimentally~\cite{2001ApPhL..79..314G}, including more involved diode designs based on two-dimensional square-lattice photonic crystals~\cite{Wang:11} or non-symmetric photonic crystal gratings exhibiting anomalous diffraction effects~\cite{Serebryannikov:2009fk}. Interesting proposals to
achieve full non-reciprocity have recently been proposed either by dynamically inducing indirect photonic transitions~\cite{Yu:2009uo} or by exploiting other acousto-optic effects induced by ultrahigh frequency
coherent acoustic waves~\cite{Kang2011}. The latter two are instances in which the time-independence of the optical response is broken due to the presence of the phonon field. However, most of these reported schemes to date require complex structures, demanding operating thresholds often with
fairly low non-reciprocal output. Most importantly they exhibit a limited tuning range, which clearly restrains the efficiency of such schemes.
\par
As dynamic optical isolation is in great demand in advanced optical communications, the question we will address here is whether a different physical principle can be used to devise a scheme that provides all-optical control over non-reciprocal transmission. An efficient way to create optical isolation is via time-reversal symmetry breaking~\cite{Wang:2009fk,fu:041112}.  Within such a regime several recent schemes have demonstrated good properties.  However, these systems exhibit little room for tuning, and either large losses~\cite{Fan}, or magnetic fields that could hamper the performance of nearby devices~\cite{Bi:2011fk}.  In contrast, we propose a mechanism where a large and tunable
non-reciprocal transmission effect can be observed in a linear, dispersive and moving Bragg multilayer.  The motion of the multilayer breaks the time-reversal symmetry.  This is used to control non-reciprocity, which may be further enhanced by optically controlling the degree of dispersion of the multilayer. Bringing together motion and control over optical dispersion constitutes a new physical approach for non-reciprocal transmission management in multilayer photonic structures.
%
%
\begin{figure}[p]
\begin{center}
\includegraphics[scale=0.5]{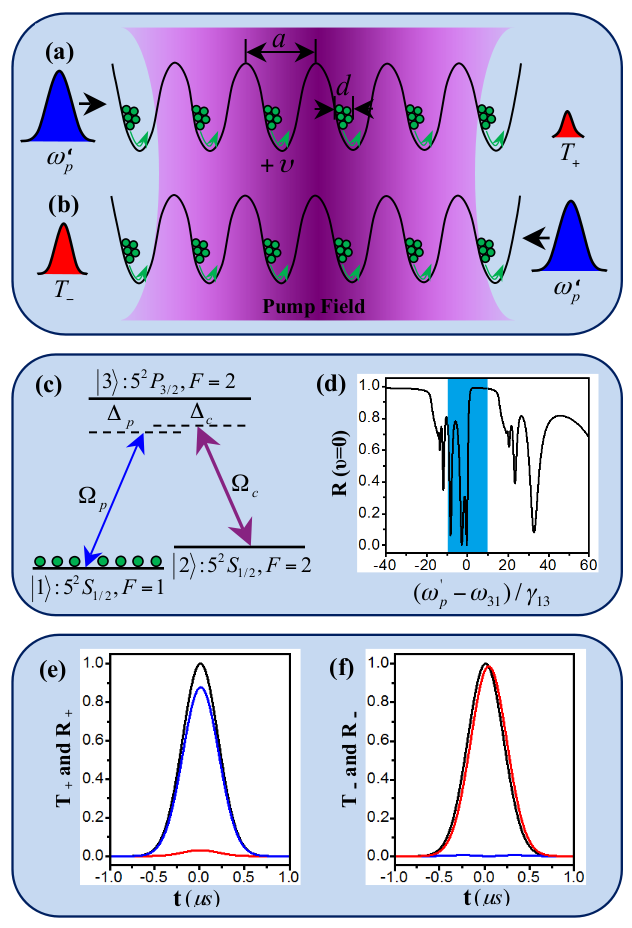}
\end{center}
\caption{\textbf{Transmission non-reciprocity in a moving atomic Bragg
mirror.} 
Cold atoms loaded into a far detuned optical lattice (\textit{dipole trap}) may arrange themselves into a 1D chain of pancake-shaped atomic layers~\protect\cite{PhysRevLett.106.223903}, of average thickness $d$ and located at the lattice antinodes (period $a$). This ordered structure, when set to move coherently~\protect\cite{Raithel:1998fk,1997PhRvL..78.2096G} inside all wells of the optical lattice, becomes a highly dispersive moving Bragg multilayer. The non-reciprocity parameter of such a structure, $\Delta T_{p}=T_{+}-T_{-}$ (see text), where $T_{+}$ corresponds \textbf{(a)} to the transmission of a light pulse moving in the forward direction with respect to the atoms velocity $\protect\upsilon$, and $T_{-}$ corresponds \textbf{(b)} to light transmission for backward incidence. \textbf{(c)} Three-level atoms used to create the Bragg mirror. The states $\left\vert 1\right\rangle $, $\left\vert 2\right\rangle $, and $\left\vert 3\right\rangle $ correspond here to the \textit{D2} line of $^{87}$Rb.  The atoms are driven into a typical $\Lambda $ configuration by the incident pulse ($\Omega _{p}$) and a pump ($\Omega _{c}$) that is used to enhance non-reciprocity~(see text). \textbf{(d)} Typical reflectivity profile of an atomic Bragg mirror at rest, as computed from Eq.(2) and plotted as a function of the scaled probe frequency detuning.  The sharp edge of a third stop-band is highlighted over the typical frequency range where the incident pulse exhibits a large nonreciprocal effect. Transmittivity (\textit{red}) and reflectivity (\textit{blue}) of a reference pulse (\textit{black}) incident from \textbf{(e)} the left and from \textbf{(f)} the right, showing a large asymmetry in the forward-backward propagation.  Sample and pulse parameters in ($\textbf{e}$)-($\textbf{f}$) are the same as in Fig.~\ref{figure_2}.
}
\label{figure_1}
\end{figure}
\par
To start with, we consider light propagating at normal incidence on a planar inhomogeneous multilayer made of isotropic, linear, non-magnetic materials with non-negligible losses, placed in an otherwise empty space and moving perpendicularly to the multilayer-vacuum interface (i.e., along the optical axis of the system). This provides the simplest example of forward/backward transmission asymmetry induced by macroscopic motion. While in the multilayer rest frame the frequency dependent transmission ${\mathcal{T}}(\omega )$ is reciprocal, in the lab frame at a given incident frequency $\omega ^{\prime}$, the transmission ${\mathcal{T}}(\omega ^{\prime })$ is in general different in the two propagation directions since the Doppler effect gives rise to two different rest frame frequencies. Specifically, consider a multilayer moving with constant velocity $\bf{\upsilon}=\upsilon \hat{\boldsymbol{x}}$ ($\upsilon /c\ll 1$) along its optical axis $x$ (see Fig.1). The frequency $\omega ^{\prime }$ of a monochromatic light beam propagating along $\pm \hat{\boldsymbol{x}}$ in the lab frame ($primed$) is seen in the rest frame ($unprimed$) as $\omega _{\pm }\simeq \left(1\mp \upsilon /c\right) \,\omega ^{\prime }$ (first order in $\upsilon /c$).  More generally for a Gaussian pulse of central frequency $\omega_{p}^{\prime }$ and spatial length ${\mathcal{L}}^{\prime }$ propagating forward ($+\hat{\boldsymbol{x}}$) and backward ($-\hat{\boldsymbol{x}}$) across the moving multilayer, the corresponding lab frame transmission~\cite{trans-pulse} can be written as,
\begin{equation}
T_{\pm }(\omega _{p}^{\prime },{\mathcal{L}}^{\prime })\simeq \frac{{%
\mathcal{L^{\prime }}}}{c\sqrt{2\pi }}\,\int_{-\infty }^{\infty }\,d\omega
^{\prime }e^{-\frac{{(\omega ^{\prime }-\omega _{p}^{\prime })}^{2}}{2(c/%
\mathcal{L^{\prime }})^{2}}}\,{\mathcal{T}}\left( (1\mp \frac{\upsilon }{c}%
)\,\omega ^{\prime }\right)  \label{TP}
\end{equation}%
Thus, forward/backward asymmetry in pulse transmission can be characterized by the non-reciprocity parameter $\Delta T_p(\omega _{p}^{\prime },{\mathcal{L^{\prime }}})\equiv T_{+}-T_{-}$. A  physical insight into the mechanism we propose may already be provided by considering a multilayer having a smooth frequency dependence of the transmissivity over the pulse bandwidth $c/\mathcal{L}^{\prime }$.  Expanding ${\mathcal{T}}(\omega )$ in a Taylor series one has~\cite{steep} $\Delta T_{p}(\omega _{p}^{\prime },{\mathcal{L^{\prime }}})\approx -\frac{\upsilon }{c}\left( 2\omega_{p}^{\prime }\frac{d{\mathcal{T}(\omega _{p}^{\prime })}}{d\omega }+(\frac{2c}{\mathcal{L}^{\prime }})^{2}\frac{d^{2}{\mathcal{T}(\omega _{p}^{\prime })}}{d\omega ^{2}}\right) $, which shows that for a nearly monochromatic pulse ($\mathcal{L}^{\prime }\rightarrow \infty $) an appreciable degree of non-reciprocity would require a steep frequency derivative of ${\mathcal{T}}(\omega )$. However, values not far from $|d{\mathcal{T}}(\omega )/d\omega|\sim c/(\omega _{p}v)$ are hardly achievable via standard multilayer systems such as \textit{e.g.} solid state dielectric Bragg mirrors~\cite{Karrai:2008fk} where non-reciprocity values of $|\Delta T_p|$ are expected to be far from unity.
\par
Such a small values of $|\Delta T_p|$ can be increased by using ultra-cold atoms in optical lattices~\cite{qusimulation}, which have been the playground over the years for the study of complex optical phenomena~\cite{citeulike:2642788,RevModPhys.77.633}.  In the following we will consider an atomic photonic structure (see Fig.1) whose constituent materials can be arranged so as to form a periodic
multilayer structure that effectively behaves as a Bragg mirror~\cite{PhysRevLett.106.223903}, and which can be set into motion at speeds of several meters per second~\cite{1997PhRvL..78.2096G, Raithel:1998fk}. Such an atomic Bragg mirror naturally exhibits resonant absorption and dispersion and represents an interesting challenge within the context of cold atoms opto-mechanics~\cite{2011PhRvL.107d3602H,2011PhRvL.107v3001C,2011PhRvA..84e1801G}.  Such a system can be realised through cooling atoms into the vibrational ground state of each potential well of a sufficiently long and red-detuned confining 1D optical lattice (\textit{dipole trap}).  The resulting atomic distribution may be described as an array of disks separated by vacuum (see Fig. 1) whose thickness $d$ is essentially given by the \textit{rms} position spread around the minima of the optical potential and is much smaller than the lattice periodicity $a$ \cite{note1}. The corresponding periodically modulated refractive index has been predicted to give rise to pronounced photonic stop-bands~\cite{PhysRevA.52.1394,coevorden484,2005PhRvE..72d6604A,PhysRevLett.96.073905}.  In spite of the fact that efficient photonic structures with cold atoms are difficult to make because dilute atomic clouds have a low refractive index contrast and intrinsically short lengths, efficient Bragg reflection has recently been demonstrated in atomic multilayers which yield a well developed one-dimensional photonic gap with nearly $80\%$ reflectivity~\cite{PhysRevLett.106.223903}.
%
%
\begin{figure}[h]
\begin{center}
\includegraphics[scale=0.6]{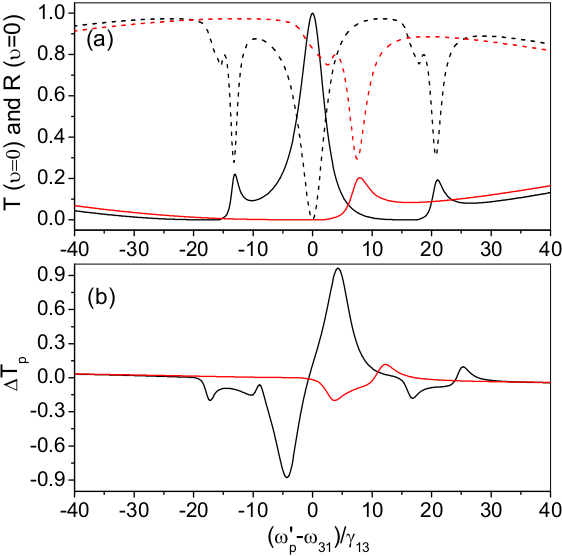}
\end{center}
\caption{Panel (a) shows the rest frame transmissivity ${\mathcal{T}}(\protect\omega )$ (solid) and reflectivity ${\mathcal{R}}(\protect\omega )$(dashed) for a dressed atomic photonic crystal with $\Omega _{c}=100$ MHz (black) or $\Omega _{c}=0.0$ (red). The sharp prominent peak of ${\mathcal{T}}$ corresponding to the edge of the third narrow stop band optically controlled by the pump (see the black curves for $\Omega _{c}=100$ MHz).  Panel (b) shows the pulse non-reciprocity parameter $\Delta T_{p}$ plotted as a function of the detuning of the probe central frequency $\protect\omega_{p}^{\prime }$ from the atomic resonance $\protect\omega _{31}$ computed for ${\mathcal{L^{\prime }}}=120$ m and $\protect\upsilon =20$ m/s with $\Omega _{c}=100$ MHz (black) or $\Omega _{c}=0.0$ (red).  Inverting the velocity would change the sign of $\Delta T_{p}$.  Other parameters are $\protect\gamma _{13}=6.0$ MHz, $\protect\gamma _{12}=1.0$ kHz, $\Delta_{c}=0.0$, $\protect\mu _{13}=1.5\times 10^{-29}$ Cm, $n=1.0\times 10^{12}$cm$^{-3}$, $l=0.5$ cm, $\protect\lambda _{31}=780.792$ nm, $\protect\lambda_{L}=780.787$ nm, $a/d=20$.}
\label{figure_2}
\end{figure}
\par
Such atomic Bragg mirrors are fairly robust against imperfections such as varying density across the mirror length and can be set to move making use of schemes that are familiar to experiments exploring the wave-packet dynamics in light-shifted
potentials~\cite{1997PhRvL..78.2096G,Raithel:1998fk}. The motion of the periodic atomic multilayers can in fact be envisaged as a collective oscillation of the atomic wave-packet in the light-shift potential wells,
much resembling that of a coherent state of atoms oscillating within a harmonic potential. Such oscillations may be induced \textit{e.g.} by suddenly shifting the optical lattice after the atoms have been
brought to equilibrium at the bottom of the lattice sites~\cite{Raithel:1998fk}. Not only can such an atomic mirror lead to unidirectional transmission effects over $95\%$, but non-reciprocity can also be effectively tuned.  Such a control (tuning) of the transmission non-reciprocity can be performed all-optically through implementation of a three-level $\Lambda $ atomic configuration in the presence of a strong pump beam (see Fig. 1).  We here refer, to be definite, to the D2 line of $^{87}$Rb atoms. The dressed atom susceptibility~\cite{RevModPhys.77.633} is then characterized by an Autler-Townes (AT) doublet rather than by a single absorption line and can be tuned on demand by changing the pump intensity and frequency as described by
\begin{equation}
\chi (\omega _{p})=\frac{n\mu _{13}^{2}}{2\varepsilon _{0}\hbar }\frac{%
\gamma _{12}-i(\Delta _{p}-\Delta _{c})}{[\gamma _{12}-i(\Delta _{p}-\Delta
_{c})](\gamma _{13}-i\Delta _{p})+\Omega _{c}^{2}}  \label{Eq2}
\end{equation}
where $\gamma _{12}$ ($\gamma _{13}$) is the dephasing rate of the spin (optical) coherence $\rho _{12}$ ($\rho _{13}$), $\Delta _{p}=\omega_{p}-\omega _{31}$ ($\Delta _{c}=\omega _{c}-\omega _{32}$) is the probe (pump) detuning relative to transition $\left\vert 1\right\rangle\leftrightarrow \left\vert 3\right\rangle $ ($\left\vert 2\right\rangle\leftrightarrow \left\vert 3\right\rangle $), while $\mu _{13}$ denotes the relevant electric-dipole moment. We assume a pump of Rabi frequency $\Omega_{c}$ travelling in the $y$ (or $z$) direction and a probe of Rabi frequency $\Omega _{p}$ travelling in the $x$ direction (the probe beam may make a small angle with the $x$ direction so to adjust the photonic crystal periodicity as in~\cite{2005PhRvE..72d6604A, PhysRevLett.106.223903}).  The dressed susceptibility in Eq.(\ref{Eq2}) holds in the limit of a weak probe ($\Omega _{p}\ll \Omega _{c}$), within the rotating-wave and electric-dipole approximations. In particular, for $\Omega _{c}\gg \gamma_{13}$, the dressed susceptibility evolves into that for two well separated AT split levels. The dressed atoms are loaded into an optical lattice of period $a$ formed by retro-reflecting a light beam of wavelength $\lambda_{L}=2a$ (see Fig. 1). As mentioned above, in each period of such a dipole trap the confined atoms occupy a small region $d\ll a$ with a homogeneous volume density $n$.
%
%
\begin{figure}[h]
\begin{center}
\includegraphics[scale=0.5]{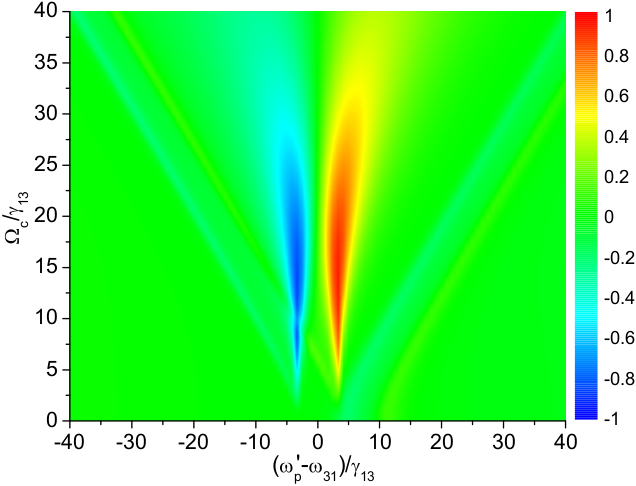}
\end{center}
\caption{Pulse non-reciprocity parameter $\Delta T_{p}$ as a function of the normalized probe detuning $(\protect\omega _{p}^{\prime }-\protect\omega_{31})/\protect\gamma _{13}$ and of the normalized pump Rabi frequency $\Omega _{c}/\protect\gamma _{13}$. Other parameters are the same as in Fig. \protect\ref{figure_2}. In particular, the pump beam has been taken to be resonant ($\Delta _{c}=0$) with the $\left\vert 2\right\rangle\leftrightarrow \left\vert 3\right\rangle $ transition.}
\label{figure_3}
\end{figure}
\par
This periodic 1D index modulation gives rise to pronounced photonic stop-bands as expected, and the transfer matrix formalism can be used to describe the propagation of a probe field of frequency $\omega $ through such an atomic stack of length $l\gg a$, to compute the transmissivity ${\mathcal{T}}(\omega )$ in its rest frame (see~\cite{2005PhRvE..72d6604A} for a detailed description in an analogous context). Notice that in a non-resonant multilayer stack, as in the case of a standard distributed Bragg reflector, a single stop-band opens up around the Bragg frequency; conversely, in the presence of a resonant absorption line close to the Bragg frequency, two stop-bands will originate from the interplay between the polaritonic gap due to the resonance and the Bragg gap due to the periodic index modulation~\cite{2005PhRvE..72d6604A}.  Yet, when the pump is on, a third narrow stop-band arises and fits between two wider ones separated by the AT splitting \cite{Petrosyan:2007uq,Schilke:2012fk}. Precisely this additional stop-band is here exploited as its edge gives rise to a very steep frequency dependence of the transmissivity ${\mathcal{T}}(\omega)$ near the atomic resonance where absorption is significantly suppressed. The width and position of this narrow stop-band depend critically on both pump frequency $\omega _{c}$ and pump intensity $\propto \left\vert \Omega_{c}\right\vert ^{2}$ [see Fig. 2(a)], which is what allows an efficient optical control of the degree of reciprocity in the transmissivity ${\mathcal{T}}(\omega)$.  Fig. 2(b) shows the non-reciprocity parameter $\Delta T_{p}$ for a light pulse incident onto a moving dressed atom photonic crystal (see Fig. 1) as a function of the pulse detuning $(\omega_{p}^{\prime }-\omega _{31})/\gamma _{13}$ of central carrier frequency from atomic resonance. A very large non-reciprocity effect appears even at a speed as slow as $\upsilon =20$ m/s owing to the steep frequency dependence of the rest frame transmissivity ${\mathcal{T}}(\omega )$, while the all optical control of $\Delta T_{p}$ is achieved through modulating the pump beam in frequency and in intensity. That is, for a $5.0$ mm long atomic lattice, the transmissivity peak reaches almost $100\%$ at the third stop-band edge and drops to near zero in less than $100$ MHz.  Correspondingly, the non-reciprocity parameter $\Delta T_{p}$ approaches $96\%$ for a nearly monochromatic pulse with the full bandwidth of $2.5$ MHz (${\mathcal{L^{\prime }}}=120$ m). In Fig. 3 we plot the non-reciprocity parameter $\Delta T_{p}$ as a function of both pump Rabi frequency $\Omega _{c}$ and probe pulse detuning $(\omega _{p}^{\prime}-\omega _{31})/\gamma _{13}$. It is clear that the best non-reciprocity effect occurs when the pump Rabi frequency is set in the AT splitting regime with $10\gamma _{13}\lesssim \Omega _{c}\lesssim 30\gamma _{13}$. Fig. 4(a) further shows the dependence of the non-reciprocity~parameter $\Delta T_{p}$ on the pulse length ${\mathcal{L^{\prime }}}$. As we can see, $\Delta T_{p}$ reduces from a value of about $92\%$ to a value of about $38\%$ for the sample velocity $\upsilon =15$ m/s when ${\mathcal{L^{\prime }}}$ is decreased from $120$ m to $10$ m. Fig. 4(b) finally shows the dependence of the non-reciprocity~parameter $\Delta T_{p}$ on the sample velocity $\upsilon $. We find that when $\Delta T_{p}$ approaches $100\%$ for ${\mathcal{L^{\prime }}}=120$ m its velocity dependence is nonlinear and tends to saturate already at $\upsilon \approx 20$ m/s.
%
%
\begin{figure}[h]
\begin{center}
\includegraphics[scale=0.6]{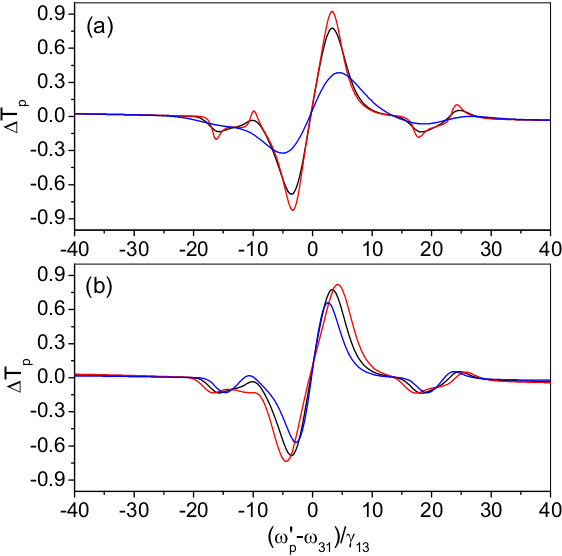}
\end{center}
\caption{Pulse non-reciprocity parameter $\Delta T_{p}$ as a function of the normalized probe detuning $(\protect\omega _{p}^{^{\prime }}-\protect\omega_{31})/\protect\gamma _{13}$ computed for $\protect\upsilon =15$ m/s and ${\mathcal{L^{\prime }}}=120$ m (red), $30$ m (black), $10$ m (blue) in (a);
for ${\mathcal{L^{\prime }}}=30$ m and $\protect\upsilon =20$ m/s (red), $15$ m/s (black), $10$ m/s (blue) in (b). Other parameters are the same as in Fig.~\ref{figure_2} (with $\Omega _{c}\equiv 100$ MHz). }
\label{figure_4}
\end{figure}
\par
It is worth noting that the wide tuning of both magnitude and sign of $\Delta T_{p}$ achieved through the present reciprocity-breaking mechanism could be exploited to realize transmission modulation in hybrid optomechanical atom-membrane interfaces where the collective motion of ultracold atoms is strongly coupled to the vibration of a micro-mechanical membrane~\cite{2011PhRvL.107v3001C, 2011PhRvA..84e1801G} (another similar means for enhancing optomechanical coupling can be found in~\cite{xuereb2012}). The unidirectional effect that we anticipate here for highly dispersive moving multilayers does
not rely on any notion that is unique to atoms and hence the scheme may also be adapted to specific optomechanical metamaterial structures where the same effect could be observed or reversibly used for mechanical sensing~\cite{Regal:2008fk,2011PhRvL.107v3001C,Aspelmeyer:2012fk}. On a more general ground, however, the reciprocity-breaking approach we suggest here is expected to stimulate further insights into the general transport theory of light in periodic structures~\cite{note2}, perhaps opening up novel possibilities in guiding and redirecting the flow of electromagnetic radiation in such media.

\acknowledgements

This work is supported by the National Natural Science Foundation of China (11174110), the National Basic Research Program of China (2011CB921603), the CRUI-British Council Programs \textquotedblleft Atoms and Nanostructures" and \textquotedblleft Metamaterials", the IT09L244H5 Azione Integrata MIUR grant, the 2011 Fondo di Ateneo of Brescia University, and the ``Malicia'' project (FET-Open grant number: 265522 of the 7th Framework Programme of the EC). Two of us (MA and GLR) would like to thank J.-H. Wu for the hospitality at Jilin University.
\bibliographystyle{naturemag}

\begin{thebibliography}{10}
\bibitem{trans-pulse}
\bibinfo{note}{$T$ is defined as the ratio of the transmitted pulse energy over
  the incident pulse energy.}
\bibitem{steep}
\bibinfo{note}{Conversely, when the transmissivity is a very steep function of
  frequency, which is the case of interest here, the pulse non-reciprocity
  parameter $\Delta T_{p}$ is to be calculated directly from (\ref{TP}).}
\bibitem{note2}
\bibinfo{note}{The scheme---as devised here---is not suitable to break reciprocity in random photonic structures, where this may be achieved e.g. through ultrafast nonlinear dephasing of light paths as shown in~\cite{muskens2012}.}
\end{thebibliography}

\end{document}